\documentclass[reprint,superscriptaddress,amsmath,amssymb,aps,pre]{revtex4-2}

\usepackage{graphicx}
\usepackage{dcolumn}
\usepackage{bm}
\usepackage{physics}
\usepackage{hyperref}
\usepackage[T1]{fontenc}
\usepackage[english]{babel}
\usepackage{float}

\begin{document}

\begin{abstract}
We propose a new approach to detailed balance violation in electrical circuits 
by relying on the scattering matrix formalism commonly used in microwave 
electronics. 
This allows to include retardation effects which are paramount at high 
frequencies.
We define the spectral densities of phase space angular momentum, heat transfer 
and cross power, which can serve as criteria for detailed balance violation. 
We confirm our theory with measurements in the 4-8 GHz frequency range on 
several two port circuits of varying symmetries, in space and time. 
This validates our approach, which allows to treat quantum circuits 
at ultra-low temperature. 
\end{abstract}

\title{Violation of detailed balance in microwave circuits: 
theory and experiment}
\author{Alexandre Dumont}
\affiliation{Département de physique and Institut Quantique, Université de 
Sherbrooke, Sherbrooke, Québec, J1K 2R1, Canada.}
\author{Pierre F\'evrier}
\author{Christian Lupien}
\affiliation{Département de physique and Institut Quantique, Université de 
Sherbrooke, Sherbrooke, Québec, J1K 2R1, Canada.}
\author{Bertrand Reulet}
\affiliation{Département de physique and Institut Quantique, Université de 
Sherbrooke, Sherbrooke, Québec, J1K 2R1, Canada.}

\date{\today}
\maketitle

\section{Introduction}
There has been recently a fast growing interest in the thermodynamics of 
ultimately simple, small systems, in particular through the study of heat 
engines~\cite{filliger_brownian_2007,argun_experimental_2017,
fogedby_autonomous_2018,sou_nonequilibrium_2019} or particles 
driven by noise sources~\cite{morgado_study_2012}, 
such as brownian motion in a fluid~\cite{martinez_brownian_2016}.
A particular effort has been devoted to electrical circuits, in which the 
variables such as position or velocity of the brownian particle are replaced by 
macroscopic variables in the circuit, such as charge or voltage across a 
capacitor, and where the noise is the Johnson-Nyquist thermal noise of 
resistors~\cite{chiang_electrical_2017,ghanta_fluctuation_2017,
gonzalez_experimental_2019}. 
All these systems, from biological entities to electrical circuits, may indeed 
obey similar equations of motion.

The simplest and most intensively studied circuit consists of two capacitors 
coupled to two resistors through another capacitor. Even such a simple circuit 
shows nontrivial heat transport~\cite{ciliberto_heat_2013,
ciliberto_statistical_2013,golubev_statistics_2015,baiesi_thermal_2016}, 
gyration and detailed balance violation~\cite{ghanta_fluctuation_2017,
gonzalez_experimental_2019}. 
Most of these studies have been performed within the framework of classical 
physics. 
On the other hand there is a currently huge development of quantum technologies 
and circuits, understanding the thermodynamical properties of which is of 
utmost interest. 
It is thus crucial to extend the methods developed in classical 
circuits to quantum ones.~\cite{bhandari_geometric_2020}

A mandatory condition to study circuits in the quantum regime is to work at 
frequencies, $f$, such that $hf\gtrsim k_BT$ with $T$ the temperature.
Since a temperature $T=1$K corresponds to a frequency $f=k_BT/h=21$ 
GHz, experiments are usually performed below 1K in the microwave domain.  
Since circuits are usually larger than the wavelength, retardation effects are 
paramount. Unfortunately, previous studies in classical circuits have focused to 
low frequencies where retardation effects are irrelevant and were not considered. 
It is the goal of the present paper to provide a new theoretical approach, 
based on the scattering matrix formalism, able to treat the case of circuits of 
any size and to validate the theory with experiments in the microwave regime.

The rest of the paper will be structured as follows: section~\ref{sec:II} is 
the theory, where we introduce the scattering matrix formalism and express the 
metrics for detailed balance violation in terms of it. 
Section~\ref{sec:III} goes over the experimental setup used to test our 
theoretical predictions in the microwave regime.
Section~\ref{sec:IV} presents the experimental data while we conclude in 
section~\ref{sec:V}.

\section{Theory}\label{sec:II}

Detailed balance refers to the absence of probability currents in phase space. It can be demonstrated using global metrics such as heat current or angular momentum in phase space.
Previous work has focused on circuits without propagation, modeling them using 
time-domain differential equations between current and voltages in the circuit, 
and where noise sources appear as source 
terms~\cite{ghanta_fluctuation_2017,gonzalez_experimental_2019}.
Below we derive expressions for these quantities when the system is described by a scattering matrix in frequency domain.

\subsection{Scattering matrix formalism}
 
We consider linear circuits where current and voltages are simply related by a 
frequency-dependent impedance matrix. Measuring voltages (respectively currents) 
requires high impedance (respectively low impedance) sensors, which are 
difficult to implement at high frequency. 
One would rather work with matched amplifiers, i.e.\ amplifiers which input 
impedance is the same as that of the transmission line connected to it, so that 
all the power sent to the amplifier is absorbed. 
Such amplifiers do not measure the voltage at a point in the circuit but the 
amplitude of the wave incoming to it. We will focus on matched amplifiers and 
discuss briefly the case of voltmeters.
We chose to work, as usual in microwave electronics, with the scattering 
formalism~\cite{pozar_microwave_2011}. 
In this formalism, a circuit connected to $n$ ports is modeled by a 
$n\cross n$ matrix, the scattering matrix $S$, which relates the amplitude of 
the voltage waves exiting each port to that entering each port. 
Sources appear as incoming waves and measurements can be performed on each 
outgoing wave.

\begin{figure}[H]
		\centering
		\includegraphics[width=1\columnwidth]{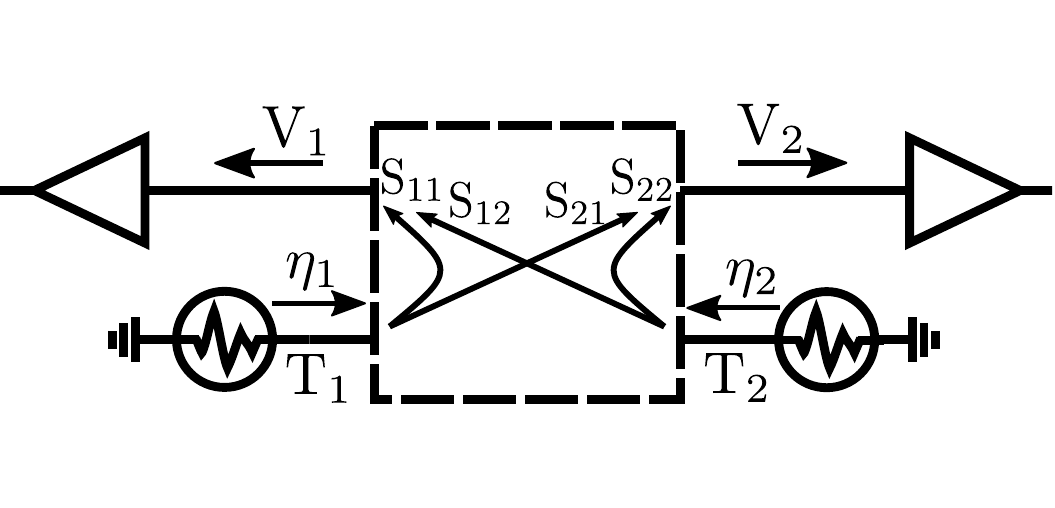}
		\caption{\textbf{(a)} Two matched resistors of value $R_0$ 
		at temperatures $T_1$ and $T_2$ and emitting voltage noises $\eta_1$ 
		and $\eta_2$ respectively.
		Both noises act as inputs into a 2 port circuit represented by its 
		four scattering parameters, while $V_1$ and $V_2$ correspond to the 
		measured outputs.}
		\label{fig:theoretical_setup}
\end{figure}

In the following we will apply the scattering formalism to the simplest, yet 
nontrivial circuit, which contains only two ports, i.e.\ two noise sources and 
two measurements, as shown in Fig.~\ref{fig:theoretical_setup}. 
This situation is very close to the one usually considered at low 
frequency~\cite{ghanta_fluctuation_2017,
gonzalez_experimental_2019,ciliberto_heat_2013,ciliberto_statistical_2013}.
This is the setup we have implemented experimentally, as we show below. 
More complex circuits can be easily treated following the same lines. 
We suppose that the circuit is lossless and linear. Losses can be 
implemented by adding extra ports in which power exits the circuit. 
According to the fluctuation-dissipation theorem, losses must be accompanied 
by noise, i.e.\ extra noise sources must be added accordingly, which enter the 
circuit through the extra ports. 
Non-linearities would complexify the physics a lot, since different frequencies 
would be coupled, and the usual scattering formalism does not apply to such 
systems. 
We do not consider non-linearities, so voltage measured at a given frequency 
depends only on noise sources at the same frequency. 
As a consequence, we can deal with the spectral densities of the various 
quantities we are considering, and not necessarily their integral over a certain 
bandwidth.

Following Fig.~\ref{fig:theoretical_setup} we note $\eta_1(f)$ and $\eta_2(f)$ 
the voltage amplitude of the waves emitted by the noise sources at frequency 
$f$, and $V_1(f)$, $V_2(f)$ the measured amplitudes of the waves leaving the 
circuit. 
They are related by:
\begin{align}
		V_1(f) &= S_{11}(f)\eta_1(f)+S_{12}(f)\eta_2(f),\label{eq:v1}\\
		V_2(f) &= S_{21}(f)\eta_1(f)+S_{22}(f)\eta_2(f)\label{eq:v2},
\end{align}
where $S_{ij}$ are the frequency-dependent elements of the $S$ matrix. For the 
sake of simplicity we suppose that the noise sources are uncorrelated and of 
spectral density $\expval{\abs{\eta_i(f)}^2}=k_BT_iR_0$. Here $R_0$ is the 
impedance of the sources, which are matched to the transmission lines connecting 
to the two ports of the circuit 
(for simplicity we take the same $R_0$ for both sources). $T_i$ is the 
(possibly frequency-dependent) noise temperature of the source $i$.
If the noise sources are resistors in the classical regime, $T_i$ is simply 
their thermodynamic temperature~\cite{johnson_thermal_1928,nyquist_thermal_1928}. If they are resistors in the quantum regime, the noise temperature $T_i$ is related to the thermodynamical temperature 
$\cal{T}$ by $T_i(f)=(hf/2k_B)\mathrm{coth}(hf/2k_B\cal{T}))$.

Below we focus on two questions: i) how to compute interesting physical 
quantities introduced in previous work, such as heat transfers and fluctuations 
loops~\cite{ghanta_fluctuation_2017}, using our approach? 
ii) can one find a better way to determine if the circuit is out of equilibrium? 

\subsection{Heat transfers}
A lot of work has been devoted to the heat transfer between two capacitively 
coupled resistors~\cite{ciliberto_heat_2013,ciliberto_statistical_2013,
golubev_statistics_2015,baiesi_thermal_2016}. 
Similar quantities can be calculated using the scattering formalism. 
Following~\cite{ciliberto_heat_2013} we note $\dot{Q}_1$ the electrical power 
dissipated in the resistor at port 1. 
Since this quantity is nonlinear in voltage, it mixes frequencies: its spectral 
density involves a convolution in frequency space. 
However the spectral density of average power $\expval{\dot{q}_1(f)}$ is well 
defined, given by:
\begin{align}
		\expval{\dot{q}_1(f)} &= \frac{1}{R_0}
		\qty[\expval{\abs{V_1(f)}^2}-\expval{\abs{\eta_1(f)}^2}]
		\label{eq:q1_dot}
\end{align}
This has a clear interpretation: it corresponds to cooling the resistor by 
emission of a wave of amplitude $\eta_1$ and heating by the absorption of a 
wave of amplitude $V_1$. 
It corresponds to the net power transfer from the left part of Fig.\ref{fig:theoretical_setup} into the circuit. Introducing the detected power spectral densities $p_i=\expval{\abs{V_i}^2}/R_0$ and their difference by $\Delta p=p_1-p_2$, we find: 
\begin{align}
		\expval{\dot{q}_1} &= \frac{\abs{S_{12}}^2}{2\abs{S_{12}}^2-1}\Delta p.
\label{eq:P_theo}
\end{align}
This result is a generalization of what has been obtained at low frequency, see Eq.\ (16) of ref.~\cite{chiang_electrical_2017}. Thus $\Delta p$ is a measure of the heat current, which vanishes at equilibrium. This also means that for any circuit, there cannot be a difference between the two power spectral densities detected unless $\Delta T=T_1-T_2\neq0$, provided that $\abs{S_{12}}^2\neq 1/2$, i.e. the circuit must not divide power equally, in which case $\Delta p$ always vanishes.

\subsection{Angular momentum and stochastic area}

It was demonstrated in~\cite{ghanta_fluctuation_2017,gonzalez_experimental_2019} 
that fluctuation loops are observed in out of equilibrium circuits. 
These loops are closed trajectories in the ($V_1$,$V_2$) plane, characterized by 
a stochastic area $A$. 
The existence of fluctuation loops corresponds to an average rotation of 
($V_1$, $V_2$) to which is associated an angular momentum along the 
perpendicular axis given by:
\begin{align}
		\expval{L_z}= \expval{V_1(t)\dot{V}_2(t)-V_2(t)\dot{V}_1(t)}
\end{align}
It is simply related to the stochastic area  by 
$\expval{L_z} = 2{\expval{\dot{A}}}$. 
This reads in Fourier space:
\begin{align}
		\expval{L_z}&=\int_{-\infty}^{+\infty} \expval{l_z(f)} df.
\end{align}
with $\expval{l_z(f)}$ the angular momentum spectral density. Experimentally the integral will have finite bounds due to the finite bandwidth of the circuit. We find:
\begin{align}
		\expval{l_z(f)} &= 4\pi f \Im\qty[\expval{V_1(f)V_2^{*}(f)}].
\label{eq:L_theo}
\end{align}
Thus the cross-correlation between $V_1$ and $V_2$ is a measure of the angular momentum, which  vanishes at equilibrium. Indeed, we find: $\expval{V_1(f)V_2^{*}(f)}/R_0=\beta k_B\Delta T$ with $\beta=S_{11} S_{21}^{*}$. This means that for any circuit, there cannot be any correlation between the two voltages detected, unless $\Delta T\neq0$. The only condition is $\beta\neq0$, i.e. the circuit must have a finite transmission. The condition on the angular momentum is however more stringent because of the imaginary part: if $\beta$ is real then $\expval{l_z}=0$. This remark sheds light on the simple reason why there are loops: the circuit must introduce a phase difference between the two branches, so that for each frequency, the two noise sources generate a rotating point in the $(V_1,V_2)$ plane. Because of the unitarity of the $S$ matrix the two rotate in opposite directions, and if the amplitude of the two noise sources are equal, there is no global rotation. The overall direction of rotation depends on the sign of $\Delta T$ as well as the sign of the phases in the $S$ matrix.

\subsection{Cross Power}
The difference in auto-correlations and the cross-correlation of the detected voltage can be used to detect deviation from equilibrium. Experimentally these two quantities can be affected by imperfections: the first is sensitive to asymmetries in the detection (amplitude mismatch) and amplifier noise, while the second is sensitive to to phase mismatch due e.g. to difference in cable lengths. The angular momentum appears to 
be related to the cross-correlation $\expval{V_1V_2^*}$ via one of it's quadrature, 
and with a weighing factor $f$. It is clear that neither the absolute 
phase of $\expval{V_1V_2^*}$ nor the frequency-dependent weighing factor are 
essential to determine if the circuit is out of equilibrium. Thus on a practical point of view it might be interesting to define the cross correlation power spectral density as
\begin{align}
		\expval{p_{1,2}} &= \frac{1}{R_0}\abs{\expval{V_1V_2^*}},
\end{align}
where we take the modulus of the cross correlation to remove the phase problem. This quantity is also a good metric of $\Delta T$ that is immune to amplifier noise and cable lengths. However it does not allow to know which source is hotter than the other.

\section{Experimental Setup}\label{sec:III}

\begin{figure}[H]
		\centering
		\includegraphics[width=0.95\columnwidth]{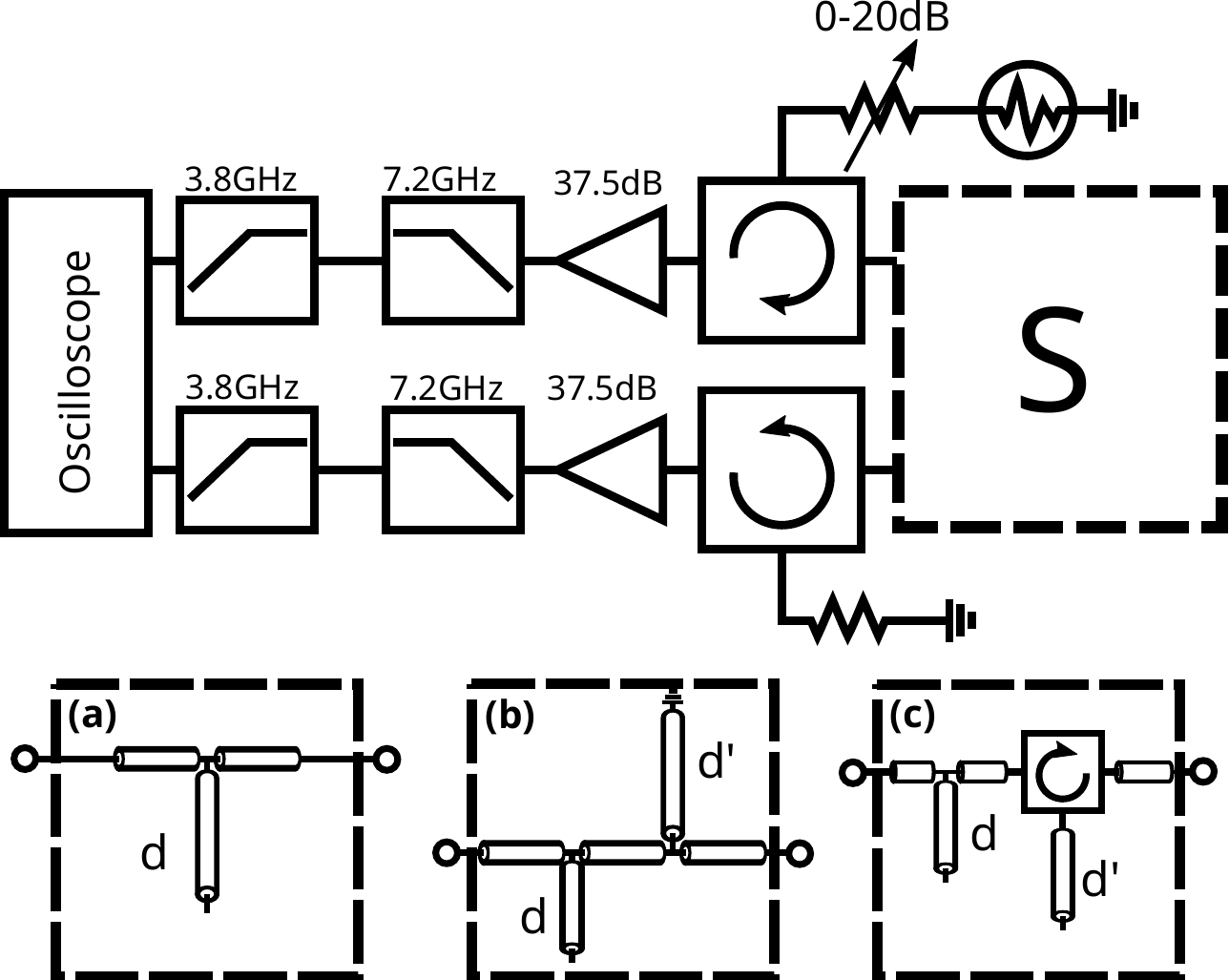}
		\caption{(Top) Experimental setup.
		The dashed box represents the coupling circuit of scattering matrix S.
		(Bottom) All three coupling circuits used to test different S matrices.}
		\label{fig:montage_general}
\end{figure}

We have performed a thorough experimental test of our theoretical results.  Beyond checking the formulas, the goal was to demonstrate the following prediction that emerges from our calculations: breaking spatial and/or time reversal symmetries cannot generate heat current or angular momentum, only 
$\Delta T$ matters. For this we performed measurements on circuits with various symmetries, see bottom of Fig.\ref{fig:montage_general}. Note that while spatial symmetry has been already considered in previous work, time-reversal symmetry could not be probed since propagation times were neglected. 
\begin{figure*}
	\centering
	\includegraphics[width=2.05\columnwidth]{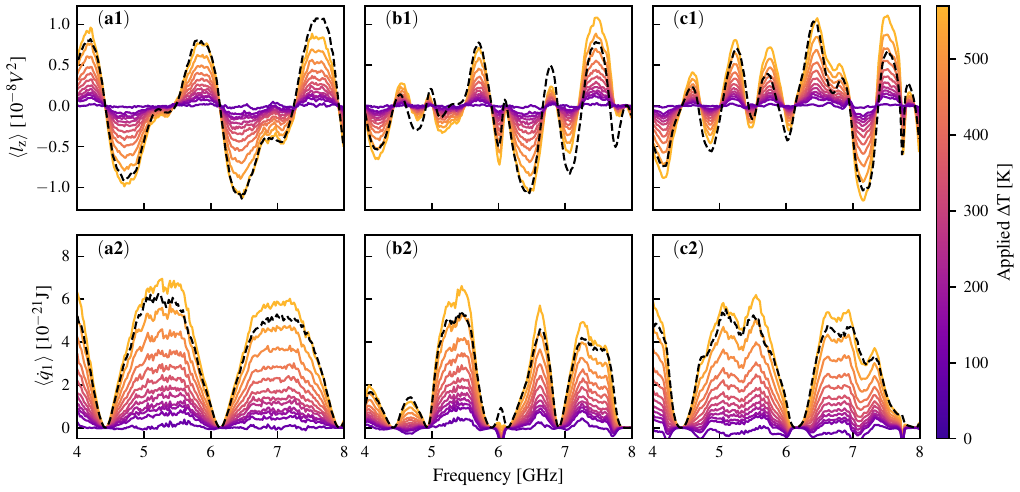}
	\caption{(a1) [resp. (b1), (c1)]: Measured average spectral density of 
	angular momentum for the circuit of Fig.~\ref{fig:montage_general}(a) 
	[resp. (b), (c)] as a function of frequency in the 4-8 GHz range. 
	Different colors correspond to different temperature differences according 
	to the color bar on the right. 
	The dashed lines represent the prediction of Eq.(\ref{eq:L_theo}) using the 
	measured scattering parameters and $\Delta T=570$K. 
	It is thus proportional to $f\Im\qty[S_{12}S_{22}^*]$.
    (a2) [resp. (b2), (c2)]: Measured average power spectral density for the 
	circuit of Fig.~\ref{fig:montage_general}(a) [resp. (b), (c)] as a 
	function of frequency in the 4-8 GHz range. Different colors correspond to 
	different temperature differences according to the color bar on the right.
	The dashed lines represent the prediction of Eq.(\ref{eq:P_theo}) using the 
	measured scattering parameters and $\Delta T=570$K. It is thus proportional 
	to $\abs{S_{12}}^2$.
	\label{fig:spectres}}
\end{figure*}

The experimental setup is shown on Fig.~\ref{fig:montage_general}.
All measurements have been performed at room temperature using a variable 
attenuator and a calibrated noise source as the hot source 
($T_1$ adjustable between 290K and 560K) and a 50$\Omega$ resistor at room 
temperature as the cold source ($T_2=290$K). 
We have chosen to work in the 4-8 GHz frequency range, which is 
similar to that of many experiments performed in the quantum regime at 
ultra low temperature. 
The separation between incoming and outgoing waves is achieved using two 
circulators: the signals emitted by the sources are injected in the circuit and 
not in the related amplifiers, while those leaving the circuits enter the 
amplifiers and are not lost in the sources. 
Moreover, the noise emitted by the amplifiers is absorbed by the sources which 
are matched to the microwave circuit. 
This minimizes parasitic cross-correlations. 
Given the noise temperature $\sim70$K of the amplifiers and isolation of the 
circulators, we estimate a parasitic contribution of $\sim0.7$K 
(more circulators could be used if needed). 
After amplification and filtering to keep the signal within a well defined 
bandwidth, the signals are digitized using a 20GHz, 40GS/s, 8 bit digital 
oscilloscope. The time series are acquired in batches 
of 2MS that are split in chunks of 2048 points. Spectra are then calculated 
using a discrete Fourier transform on each chunk. 
Then we calculated the auto- and cross-correlations using those spectra and 
averaged over all the chunks. 
The size of the chunk sets the frequency resolution, here 20MHz, which is enough for the circuits we considered.

In order to test the effect of spatial/time reversal symmetries, we have studied the three circuits shown in Fig.~\ref{fig:montage_general}. 
The circuits are made of pieces of coax cables of various lengths connected by 
T-junctions, and terminated by open or short circuits.
The lengths $d,d'$ of 5.08 and 7.62 cm respectively, 
where chosen so that reflections of waves at the end of the cables and at the 
junctions provide rich interference patterns which result in strong 
frequency-dependence of the S-matrix, thus providing a deep test of our 
theoretical results, see dashed lines in Fig.~\ref{fig:spectres}. 
Here propagation times are essentials. 
Circuit (a) in Fig.~\ref{fig:montage_general} is symmetric upon exchange of 
ports 1 and 2 while circuit (b) is not. Circuit (c) contains a circulator in 
order to break time-reversal symmetry.
In order to make a quantitative comparison between our predictions and the 
measurements, we have measured the $S$ matrix of all 
three circuits using a vector network analyzer 
(VNA), see dashed lines in Fig.~\ref{fig:spectres}.

\section{Results}\label{sec:IV}
In Fig.~\ref{fig:spectres} we show the spectral densities of the angular 
momentum $\expval{l_z(f)}$ (top) and transferred power $\expval{\dot{q}_1(f)}$ 
(bottom) as a function of frequency for the three circuits of 
Fig.~\ref{fig:montage_general}. 
The two quantities exhibit strong oscillations vs.\ frequency which come from 
interferences occurring due to total reflections at the end of the cables and 
partial reflections at the junctions between them. 
We observe a very good quantitative agreement between the measurements and 
theoretical predictions of Eqs. (\ref{eq:L_theo}) and (\ref{eq:P_theo}), shown 
as dashed lines in Fig.~\ref{fig:spectres}. 
These have been obtained using the measured coefficients of the scattering 
matrix and the calibrated noise temperature of the hot noise source. 
We attribute the small differences between theory and experiment to experimental 
imperfections, in particular the lack of reproducibility of 
connections/disconnections between the VNA and noise measurements, and the 
losses in the cables and circulator.
The slightly negative values of $\langle\dot{q_1}\rangle$ are due to the presence of amplifier 
noise which is not accounted for in Eq.~(\ref{eq:P_theo}).

\begin{figure}[H]
		\centering
		\includegraphics[width=1\columnwidth]{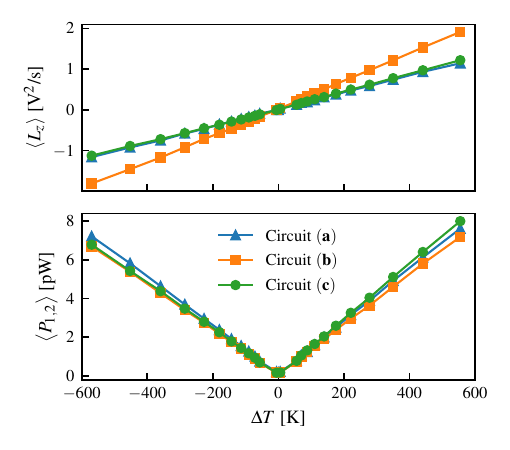}
		\caption{(Top) Total angular momentum in the frequency range 
		$7.3 - 7.5$ GHz for the three circuits of 
		Fig.~\ref{fig:montage_general} as a function of $\Delta T$. 
		(Bottom) Total cross-power in the frequency range
		$4-8$ GHz for the three circuits of 
		Fig.~\ref{fig:montage_general} as a function of $\Delta T$.}
		\label{fig:linear}
\end{figure}

The dependence of the integrated angular momentum $\expval{L_z}$ and 
cross-power $\expval{P_{1,2}}$ on the 
temperature difference $\Delta T$ is shown in Fig.~\ref{fig:linear}. 
These curves have been obtained by integrating the spectra of 
Fig.~\ref{fig:spectres} over frequency between 7.3 and 7.5 GHz for the angular 
momentum and the full 4-8 GHz bandwidth for the cross power (the relatively narrow bandwidth for $\expval{l_z(f)}$ has been chosen to 
avoid sign change leading to a vanishing $\expval{L_z}$). 
The data corresponding to $\Delta T<0$ have been obtained by 
swapping the cold and hot sources.
As predicted, $\expval{L_z}\propto\Delta T$, and 
$\expval{P_{1,2}}\propto\abs{\Delta T}$. This is true for 
all circuits, i.e. is independent of the presence or absence of spatial and/or time-reversal symmetries 
in the system.

We notice that $\expval{P_{1,2}}$, which in theory should be proportional to 
$\abs{\Delta T}$ is experimentally not exactly a perfectly even function of 
$\Delta T$. 
This comes from the impedances of the sources not being exactly 50 $\Omega$, so 
reversing the sources also slightly changes $S$. 
The same thing happens to $\expval{L_z}$ which is not perfectly odd, although 
it is less obvious. 
For $\Delta T=0$ we find a residual cross-power which corresponds to what we 
expect for $\Delta T\sim13$K. We understand this as amplifier noise ($\sim70$K) 
being averaged a finite number $N=100$ of times. 
This yields $\Delta T\approx 70/\sqrt{100}=7$K, in reasonable agreement with 
the measurement.  
We also find a residual angular momentum corresponding to $\Delta T\sim0.6$K, 
in good agreement with our prediction, see section~\ref{sec:III}.

\section{Conclusion}\label{sec:V}
We have shown, both theoretically and experimentally how heat transfer and angular momentum, quantities that have been introduced in electrical circuits at low frequency in the context of detailed balance violation, can be extended to microwave circuits where propagation delays are paramount.
Our approach is very general and could be used not only for microwaves but also 
for guided optics for example. In particular, it allows to treat circuits in 
the quantum regime $\qty(hf\gtrsim k_{\text{B}}T)$ where the noise spectra of the 
sources have a frequency dependence which reflects the presence of vacuum 
fluctuations\cite{golubev_statistics_2015}.

From our result one should be able to recover previously obtained results at 
low frequency where propagation times 
vanish~\cite{ciliberto_heat_2013,ciliberto_statistical_2013}. 
For this we have to consider voltmeters and not only matched amplifiers as we 
did. 
This can be done as follows: a voltmeter does not measure the outgoing voltage 
$V_{out}$ but the sum of the incoming $V_{in}$ and outgoing voltages, 
$V_{in}+V_{out}$. 
As a consequence, our approach is valid provided we replace the scattering 
matrix $S$ by $S+1$. However, the latter matrix being non-unitary, 
simplifications of the formula cannot be carried out. 

We have focused on the detection of detailed balance violation using average quantities: average heat transfer, average angular momentum.
These involve the measurement of the second order correlations of the detected voltages. 
Some previous works have considered fluctuations of these quantities and 
calculated their probability 
distribution~\cite{ciliberto_heat_2013,ciliberto_statistical_2013,
golubev_statistics_2015}. 
With our approach, it is possible to reconstruct these distributions by measuring their moments. 
For example, the variance of the angular momentum is related to correlations of order 4 of the measured voltages.
Dealing with moments would be interesting in particular with non-gaussian noise: what happens in a circuit with two noise sources of equal variance but opposite third moments, generated for example by two shot noise sources with opposite bias ?

\section*{Acknowledgements}
The authors would like to acknowledge the many conversations with Clovis Farley 
and Simon Bolduc Beaudoin.
This work was supported by the Canada Research Chair program, 
the NSERC, the Canada First Research Excellence Fund, the FRQNT, and the Canada 
Foundation for Innovation.

\bibliography{references}

\end{document}